**LOAD FLOW ANALYSIS OF RADIAL DISTRIBUTION NETWORK USING LINEAR DATA STRUCTURE**

A

***Dissertation***

*submitted*

*in partial fulfillment*

*for the award of the Degree of*

***Master of Technology***

***in Department of Computer Science & Engineering***

**(with specialization in Computer Science)**

**Supervisor:**

Mr. Arnab Maiti
Assistant Professor

**Submitted By:**

Ritu Parasher
10E2YTCSF4XT615

**Department of Computer Science & Engineering**

**Yagyavalkya Institute of Technology, Jaipur**

**Rajasthan Technical University, Kota**

# CANDIDATE'S DECLARATION

I hereby declare that the work, which is being presented in the Dissertation, entitled "**Load Flow Analysis of Radial Distribution Network Using Linear Data Structure**" in partial fulfillment for the award of Degree of "**Master of Technology**" in **Department of Computer Science & Engineering** with Specialization in **Computer Science** and submitted to the Department of Computer Science & Engineering, **Yagyavalkya Institute of Technology**, is a record of my own investigations carried under the Guidance of **Mr. Arnab Maiti,** Department of Computer science & Engineering, Yagyavalkya Institute of Technology, Jaipur.

I have not submitted the matter presented in this Dissertation anywhere for the award of any other Degree.

**(Ritu Parasher)**
**Computer Science & Engineering,**
**Enrolment No.: 10E2YTCSF4XT615**
**Yagyavalkya Institute of Technology, Jaipur**

**Counter Signed by**

**(Mr. Arnab Maiti)**
**Assistant Professor**
**Computer Science & Engineering,**
**Yagyavalkya Institute of Technology, Jaipur**

# ACKNOWLEDGEMENT

First of all, I thank the Almighty God, who gave me the opportunity and strength to carry out this work.

I express my profound and sincere gratitude to **Dr. Pankaj Sharma**, Principal, Yagyavalkya Institute of Technology, Jaipur for providing all the facilities and support during my academic and project period.

"Expression of feelings by words makes them less significant when it comes to make statement of gratitude".

I would like to thank my guide **Mr. Arnab Maiti** for his valuable guidance, constructive criticism and encouragement and also for making the requisite guidelines to enable me to complete my dissertation work.

I would like to express my deep sense of gratitude to **Mr. Ankur Goyal,** Head of the Department (CSE&IT), Mr.Yogesh Rathi, Mr Manish Jain, Dr. Ashutosh Sharma Assistant Professors of Yagyavalkya Institute of Technology, Jaipur for their innovative ideas and great support for completing my project work.

I express my deep thanks to My Grandfather, My Parents, My Sisters, My Brothers, and all my family members for the motivation, inspiration and support in boosting my moral without which I would have been in vain.

I am also thankful to the previous researchers whose published work has been consulted and cited in my dissertation.

**(Ritu Parasher)**

# TABLE OF CONTENTS

# LIST OF FIGURES

# LIST OF TABLES

# LIST OF ABBREVATIONS

$S_i$ : Complex power fed at node i

$P_i$ : Real Power fed at node i

$Q_i$ : Reactive Power fed at node i

NB : Total number of nodes (i = 1,2, -------- NB)

LN : Total number of branches (LN =NB-1)

$PL_i$ : Real power load at $i^{th}$ node

$QL_i$ : Reactive power load at $i^{th}$ node

$|V_i|$ : Voltage magnitude of $i^{th}$ node

$\theta_{v_i}$ : Voltage angle of $i^{th}$ node

$|LI_i|$ : Load current magnitude at $i^{th}$ node

$\theta_i$ : Load current angle at $i^{th}$ node

$|I_{br_j}|$ : Current magnitude in branch j

$\angle I_{br_j}$ : Current angle in branch j

$V_s$ : Sending node voltage

$V_r$ : Receiving node voltage

IS (j) : Sending end node of branch j

IR (j) : Receiving end node of branch j

$LP_j$ : Real power loss of branch j

$LQ_j$ : Reactive power loss of branch j

RDN : Radial distribution network

LFA : Load flow analysis

# ABSTRACT

Distribution systems hold a very significant position in the power system since it is the main point of link between bulk power and consumers. A planned and effective distribution network is the key to cope up with the ever increasing demand for domestic, industrial and commercial load. The load-flow study of radial distribution network is of prime importance for effective planning of load transfers. Power companies are interested in finding the most efficient configuration for minimization of real power loses and load balancing among distribution feeders to save energy and enhance the over all performance of distribution system.

This thesis presents a fast and efficient method for load-flow analysis of radial distribution networks. The proposed method is based on linear data structure. The order of time and space complexity is reported here. There is significant saving in no. of steps execution. Using graph theory concept and exploiting multi-cores architecture, the proposed method for load flow can be further investigated for obtaining more optimized solutions.

# INTRODUCTION

## 1.1 Introduction

Distribution systems hold a very significant position in the power system since it is the main point of link between bulk power and consumers. Effective planning of radial distribution network is required to meet the present growing domestic, industrial and commercial load day by day. Distribution networks have gained an overwhelming research interest in the academics as well as in the industries community nearly from last three decades. The examples of prominent distribution networks that effect domestic/residential users and industrial personals are water distribution networks, electricity distribution networks, data/voice communication networks, and road traffic networks etc. Electricity is an essential commodity and its absence for short-while creates annoyance and discomfort in everybody's life. In fact, it puts most of the modern household and office appliances to a total stop . Electrical power distribution is either three or four wires. In these four wires,3 wire are for phases and 1 wire for Neutral. The Voltage between phase to phase is called Line Voltage and the voltage between phase and neutral is called Phase Voltage. This forth wire may or may not be distributed and in the same way this neutral may or may not be earthed.. The neutral may be directly connected to earth or connected through a resistor or a reactor. This system is called Directly earthed or Earthed system. In a network, the earthing system plays a very important role. When an insulation fault occurs or a phase is accidentally earthed, the values taken by the fault currents, the touch voltages and over voltages get closely linked to the type of neutral earthing connection. A directly earthed neutral strongly limits over voltages but it causes very high fault currents, same as an unearthed neutral limits fault currents result to very low values but encourages the occurrence of high over voltages. In any installation, service continuity in the event of an insulation fault is also directly related to the earthing system. An unearthed neutral permits service continuity during an insulation fault. Contrary to this, a directly earthed neutral, or low impedance-earthed neutral, causes tripping as soon as the first insulation fault occurs. In order to meet these specifications, a properly designed and operated radial distribution network should possess the following characteristics.

1. The system should support energy supply at minimum operation and maintenance cost and should satisfy the social and engineering aspects.
2. It must satisfy the continuous changing of the load demand for active and reactive power.
3 Unlike other forms of energy, electricity is not easily stored and thus, adequate “spinning” reserve of active and reactive power should be maintained and controlled in an appropriate manner.
4. The power supply must meet the following specific standards to maintain the quality of service offered
   a. Regulated voltage
   b Well maintained constant frequency
   c. Level of reliability/security that guarantees consumers satisfaction.

## 1.2 Objective

The objective of thesis work is to propose a new algorithm. This algorithm exploits stack data structure for branch currents computation since stack data structure is a linear data structure so this algorithm can also be called as linear data structure based Load Flow Analysis algorithm. It is key issue addressed in proposed algorithm to facilitate an easy calculation of currents in all the branches. In fact time and space complexity used by algorithm are the two main measures that decide the efficiency of algorithm.

## 1.3 Typical Power Network

An understanding of basic design principles is essential in the operation of electric power systems. This chapter briefly describes and defines electric power generation, transmission, and distribution systems (primary and secondary). A discussion of emergency and standby power systems is also presented. Figure 1.1 shows a one-line diagram of a typical electrical power generation, transmission, and distribution system.

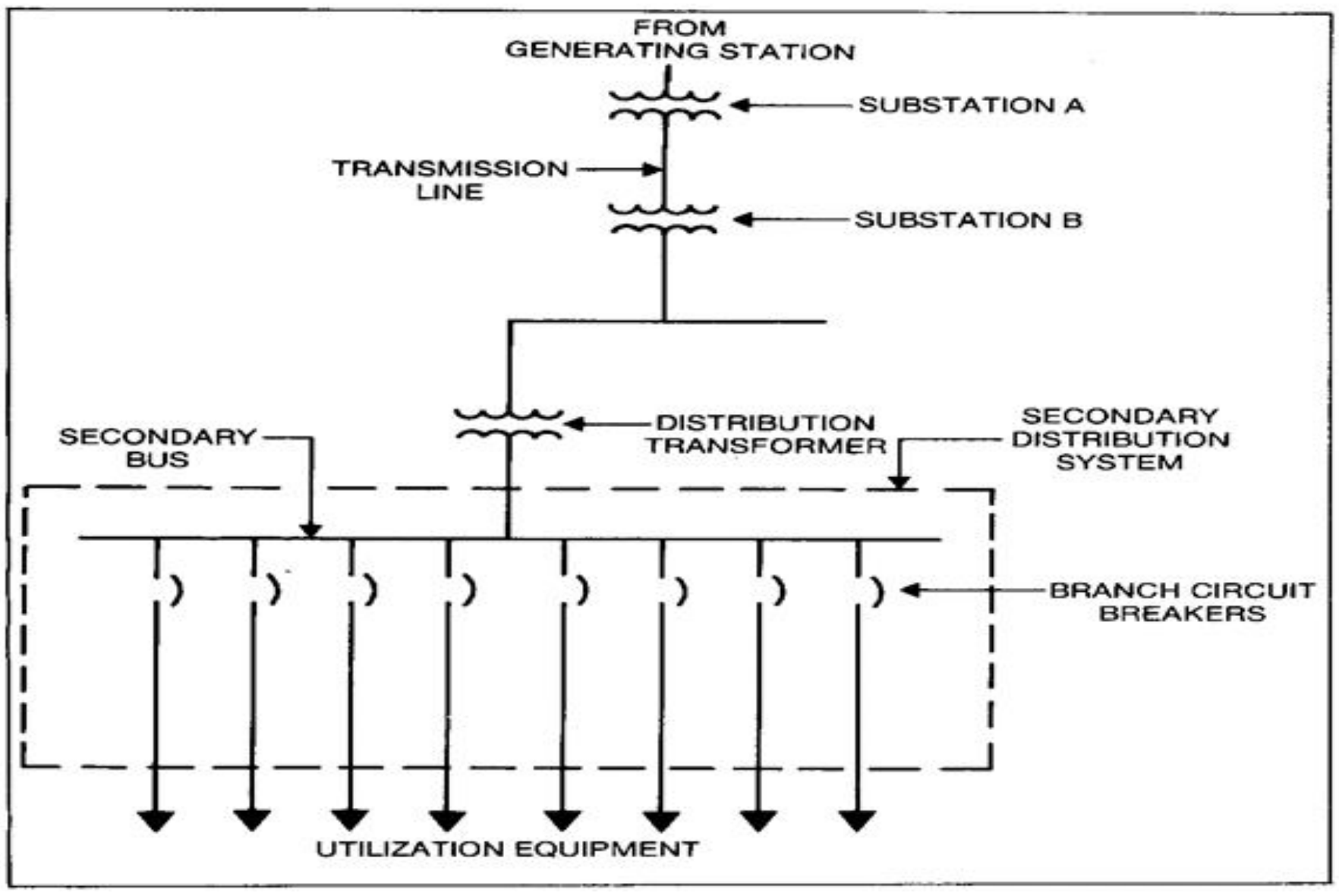

**Figure.1.1 : Typical Electric Power Generation, Translation and Distribution System**

The transmission systems are basically a bulk power transfer links between the power generating stations and the distribution sub-stations from which the power is carried to customer delivery points. The transmission system includes step-up and step-down transformers at the generating and distribution stations, respectively. The transmission system is usually part of the electric utility's network. Power transmission systems may include sub-transmission stages to supply intermediate voltage levels. Sub-transmission stages are used to enable a more practical or economical transition between transmission and distribution systems. It operates at the highest voltage levels (typically, 230 kV and above). The generator voltages are usually in the range of 11 kV to 35 kV. There are also a few transmission networks operating in the extremely high voltage class (345 kV to 765 kV). As compared to transmission system sub-transmission system transmits energy at a lower voltage level to the distribution substations. Generally, sub-transmission systems supply power directly to the industrial customers. The distribution system is the final link in the transfer of electrical energy to the individual customers. Between 30 to 40% of total investment in the electrical sector goes to distribution systems, but nevertheless, they haven’t received the technological improvement in the same manner as the generation and transmission systems. The distribution network differs from its two of siblings in topological structure as well as its associated voltage levels. The distribution networks are generally of radial or tree structure and hence referred as Radial Distribution Networks (RDNs). Its primary voltage level is typically between 4.0 to 35 kV, while the secondary distribution feeders supply residential and commercial customers at 120/240/440 volts. In general, the distribution system is the electrical system between the substation fed by the transmission system and the consumers' premises. It generally consists of feeders, laterals (circuit-breakers) and the service mains.

## 1.4 Elements of the Distribution System

In general, the distribution system is derived from electrical system which is Substationally fed by the consumers' premises and the transmission system. It generally consists of feeders, laterals (circuit-breakers) and the service mains. Figure1.2 shows the single line diagram of a typical low tension distribution system.

### 1.4.1 Distributed Feeders

A feeder is a conductor, which connects the sub-station (or localized generating station) to the area where power is to be distributed. Generally, no tapping are taken from the feeder so that the current in it remains the same throughout. The main consideration in the design of a feeder is the current carrying capacity.

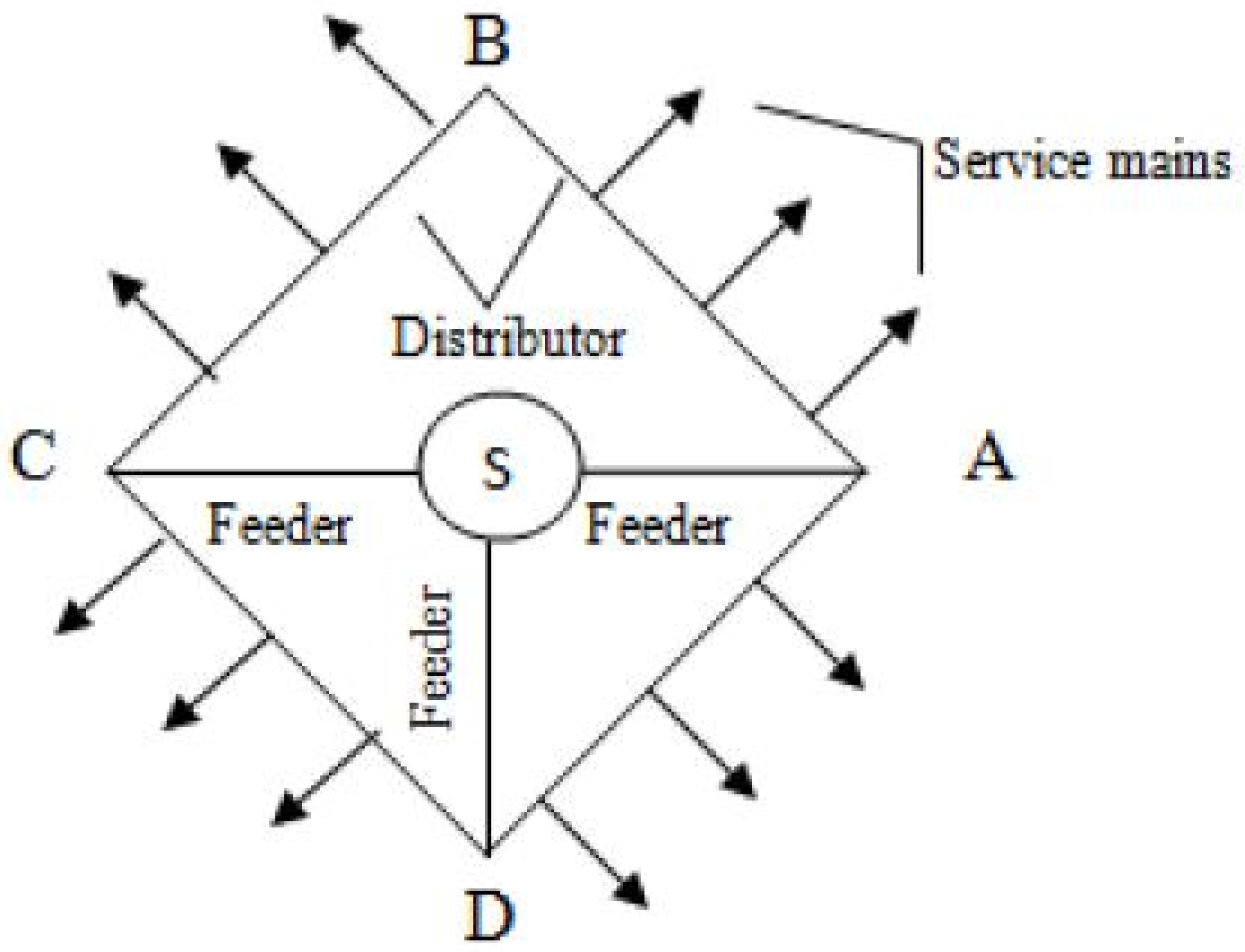

**Figure 1.2 : Elements of Distribution System**

### 1.4.2 Distributor

A distributor is a conductor from which tapping are taken for supply to the consumers. In Figure1.2, AB, BC, CD, and DA are the distributors.

The current through a distributor is not constant because tapping are taken at various places along its length. While designing a distributor, voltage drop along its length is the main consideration since the statutory limit of voltage variations is 10% of rated value at the consumer's terminals.

### 1.4.3 Service mains

A service mains is generally a small cable which connects the distributor to the consumer's terminals.

## 1.5 Requirements of a Distribution System

It is mandatory to maintain the supply of electrical power within the requirements of many types of consumers. Following are the necessary requirements of a good distribution system:

**(a)** **Availability of power demand:** Power should be made available to the consumers in large amount as per their requirement. This is very important requirement of a distribution system.

**(b)** **Reliability:** As we can see that present day industry is now totally dependent on electrical power for its operation. So, there is an urgent need of a reliable service. If by chance, there is a power failure, it should be for the minimum possible time at every cost. Improvement in reliability can be made upto a considerable extent by

a) Reliable automatic control system.

b) Providing additional reserve facilities.

**(c)** **Proper voltage**: Furthermost requirement of a distribution system is that the voltage variations at the consumer terminals should be as low as possible. The main cause of changes in voltage variation is variation of load on distribution side which has to be reduced. Thus, a distribution system is said to be only good, if it ensures that the voltage variations are within permissible limits at consumer terminals.

**(d)** **Loading** : The transmission line should never be over loaded and under loaded.

**(e)** **Efficiency** : The efficiency of transmission lines should be maximum say about 90%.

## 1.6 Classification of Distribution System

A distribution system may be classified on the basis of:-

i) Nature of current: According to nature of current, distribution system can be classified as

a) AC distribution system.

b) DC distribution system.

ii) Type of construction: According to type of construction, distribution system is classified as

a) Overhead system

b) Underground system

iii) Scheme of operation: According to scheme of operation, distribution system may be classified as:

a) Radial delivery network

b) Ring main system

c) Interconnected system

With the growing market in the present time, power flow analysis has been one of the most fundamental and an essential tool for power system operation and planning. In RDN, each of its branch or link has a unique path for power flow from the substation (source of energy) i.e. source node to end (leafs) nodes.

With the rapid development in computing techniques since the 1970s, many power flow algorithms based on modern computing techniques have been reported. According to these studies, power flow analysis of RDNs may be divided into two categories. The first group of methods includes: ladder network methods for radial structure distribution systems using basic laws of circuit theories like Kirchhoff's Current Law (KCL) and Kirchhoff's Voltage Law (KVL). On the other hand, the second category includes Gauss-Seidel, Newton-Raphson and Decoupled Newton-Raphson methods for transmission systems and are usually based on nodal analysis method. According to few of the reported studies by, Salama and Chikhani [1], Da Costa, Martins, and Pereira [2] and Srinivas [3] Shirmohammadi, Hong, Semlyen and Luo [4] following conclusions can be drawn:

- Ladder network methods implementation finds suitability and practical application for single source radial structure networks with high R/X ratio;
- Nodal analysis methods are suitable for multiple-source systems.
- The characteristics of RDNs are dynamic in nature

## 1.7 Features of RDN

1. Uncertainties and Imperfection of network parameters.
2. High R/X ratio
3. Extremely large number of nodes and branches.
4. Dynamic change in imposed load.

## 1.8 Ring main system

The loop (or ring) distribution system is one that starts at a distribution substation, runs through or around an area serving one or more distribution transformers or load centre, and returns to the same substation.

The ring main system has the following advantages:

a) There are very less voltage fluctuations at consumer’s terminals.

b) The system is very reliable as each distributor is fed with two feeders. In case, of fault in any section of feeder, the continuity of supply is maintained.

## 1.9 Organization of Dissertation Work:

The dissertation comprises of five chapters. The brief description of each chapter is as follows.

Chapter 1 starts with a brief descriptions on electric power distributions system, load flow analysis of distribution system and network reconfiguration of distribution system. It also includes the objective of the dissertation; research methodology used to serve this purpose. Chapter 2 begins with the review of the literature on load-flow analysis of radial distribution System and then scope of the research. Chapter 3 presents the load-flow analysis of radial distribution networks. The assumption, load flow modeling and algorithms. Chapter 4 presents the results. Chapter 5 presents the conclusion and the future scope. References present the list of previous papers published by researchers in load-flow, and Consist of the list of the related books that have been surveyed by the author.

# LITERATURE SURVEY

## 2.1 Introduction

Load flow studies are used to ensure that electrical power transfer from generators to consumers through the grid system is stable, reliable and economic. The increasing presence of distributed alternative energy sources, often in geographically remote locations, complicates load flow studies and has triggered a resurgence of interest in the topic. In a three phase ac power system active and reactive power flows from the generating station to the load through different networks buses and branches. The flow of active and reactive power is called power flow or load flow. Power flow studies provide as systematic mathematical approach for determination of various bus voltages, there phase angle active and reactive power flows through different branches, generators and loads under steady state condition. In order to obtain a reliable power system operation under normal balanced three phase steady state conditions, it is required to have the followings:

- Generation supplies the load demand and losses.
- Bus voltage magnitudes remain close to rated values.
- Generator operates within specific real and reactive power limits.
- Transmission lines and transformers are not overloaded.

Power flow analysis is used to determine the steady state operating condition of a power system. Power flow analysis is widely used by power distribution professional during the planning and operation of power distribution system

## 2.2 Literature Survey

In the literature, there are a number of efficient and reliable load flow solution techniques, such as; Gauss-Seidel, Newton-Raphson and Fast Decoupled Load Flow [5,6,7,8,9,10,11]. In 1967, Tinney and Hart [5] developed the classical Newton based power flow solution method. Later work by Stott and Alsac [6] made the fast decoupled Newton method. The algorithm made by [6] remains unchanged for different applications. Even though this method worked well for transmission systems, but its convergence performance is poor for most distribution systems due to their high R/X ratio which deteriorates the diagonal dominance of the Jacobian matrix. For this reason, various other types of methods have been presented. Those methods consist of backward/forward sweeps on a ladder system. The formulation of the algorithm for those methods were different from the Newton's power flow method, which made those methods hard to be extended to other applications in which the Newton method seemed more appropriate.

Tripathy et al. [12] presented a Newton like method for solving ill-conditioned power systems. Their method showed voltage convergence but could not be efficiently used for optimal power flow calculations.

Baran and Wu [13], proposed a methodology for solving the radial load flow for analyzing the optimal capacitor sizing problem. In this method, for each branch of the network three non-linear equations are written in terms of the branch power flows and bus voltages. The number of equations was subsequently reduced by using terminal conditions associated with the main feeder and its laterals, and the Newton- Raphson method is applied to this reduced set. The computational efficiency is improved by making some simplifications in the jacobian.

Chiang [14] had also proposed three different algorithms for solving radial distribution networks based on the method proposed by Baran and Wu .He had proposed decoupled, fast decoupled & very fast decoupled distribution load-flow algorithms. In fact decoupled and fast decoupled distribution load-flow algorithms proposed by Chiang [14] were similar to that of Baran and Wu [13].

Goswami and Basu [15] had presented a direct method for solving radial and meshed distribution networks. However, the main limitation of their method is that no node in the network is the junction of more than three branches, i.e one incoming and two outgoing branches.

Jasmon and Lee [16] had proposed a new load-flow method for obtaining the solution of radial distribution networks. They have used the three fundamental equations representing real power, reactive power and voltage magnitude derived in [15]. They have solved the radial distribution network using these three equations by reducing the whole network into a single equivalent.

Das et al. [17] had proposed a load-flow technique for solving radial distribution networks by calculating the total real and reactive power fed through any node using power convergence with the help of coding at the lateral and sub lateral nodes for large system that increased complexity of computation. This method worked only for sequential branch and node numbering scheme. They had calculated voltage of each receiving end node using forward sweep. They had taken the initial guess of zero initial power loss to solve radial distribution networks. It can solve the simple algebraic recursive expression of voltage magnitude and all the data can be easily stored in vector form, thus saving an enormous amount of computer memory.

Haque [18] presented a new and efficient method for solving both radial and meshed networks with more than one feeding node. The method first converted the multiple-source mesh network into an equivalent single-source radial type network by setting dummy nodes. Then the traditional ladder network method could be applied for the equivalent radial system. Unlike other method effect of shunt and load admittances are incorporated in this method because of which it can be employed to solve special transmission networks. This method has excellent convergence for radial network.

Eminoglu and Hocaoglu [19] presented a simple and efficient method to solve the power flow problem in radial distribution systems which took into account voltage dependency of static loads, and line charging capacitance. The method was based on the forward and backward voltage updating by using polynomial voltage equation for each branch and backward ladder equation. The proposed power flow algorithm has robust convergence ability when compared with the improved version of the classical forward-backward ladder method, i.e., Ratio-Flow.

Prasad et. al. [20] proposed a simple and efficient scheme for computation of the branch currents in RDN. The proposed load flow algorithm exploits the tree- structure property of a RDN and claims the efficient implementation of LFA algorithm. However, in the reported work authors have exercised the leaf nodes identification in each iteration of LFA algorithm till voltage estimation satisfies the stated convergence criteria. This procedure of leaf node identification leads to an overburden LFA algorithm for static as well as dynamic network topologies.

Ghosh and Sherpa [21] presented a method for load-flow solution of radial distribution networks with minimum data preparation. This method used the simple equation to compute the voltage magnitude and has the capability to handle composite load modeling. But in order to implement this algorithm, a lot of programming efforts are required.

Sivanagaraju et al. [22] proposed a distinctive load flow solution technique which is used for the analysis of weakly meshed distribution systems. A branch-injection to branch-current matrix is formed (BIBC) and this matrix is formed by applying Kirchhoff's current law for the distribution network. Using the same matrix that is BIBC a solution for weakly meshed distribution network is proposed.

Kumar and Arvindhababu [23] presented an approach for power flow solutions to obtain a reliable convergence in distribution systems. The trigonometric terms were eliminated in the node power expressions and thereby the resulting equations were partially linearized for obtaining better convergence. The method was simpler than existing approaches and solved iteratively similar to Newton-Raphson (NR) technique.

Augugliaro et al. [24] had proposed a method for the analysis of radial or weakly meshed distribution systems supplying voltage dependent loads. The solution process is iterative and at every step loads are simulated by impedances. Therefore it is necessary to solve a network made up only of impedances; for radial systems, all the voltages and currents are expressed as linear functions of a single unknown current and for mesh system two unknown currents for each independent mesh. Advantages of this method are: its possibility to take into account of any dependency of the loads on the voltage, very reduced computational requirements and high precision of results.

Gurpreet kaur [25] In this thesis, a new method of load-flow technique for solving radial distribution networks by sequential numbering scheme has been proposed. The aim of this thesis is to reduce data preparation and propose a method to identify the nodes beyond each branch with less computation.

D.P Sharma [26] in this thesis, two new efficient load flow algorithms along with couple of new schemes for network reconfigurations are investigated and simulation test results are presented.

## 2.3 Scope of the Research

Literature survey shows that a number of methods had been proposed for load-flow solution of radial distribution networks. In some cases authors had used the data as it is without reducing data preparation and in some cases authors have tried to reduce the data preparation. Since the distribution system is radial in nature having high R/X ratio, the load flow methods become complicated. The aim of this thesis work is to reduce the data preparation using the sequential numbering scheme and the radial feature of distribution networks. The proposed method not only reduces the data preparation but also increases the efficiency of the load flow.

# LOAD FLOW ANALYSIS OF RADIAL DISTRIBUTION NETWORK

## 3.1 Introduction

A planned and effective distribution network is the key to cope up with the ever increasing demand for domestic, industrial and commercial load. The load-flow study of radial distribution network is of prime importance for effective planning of load transfer. In majority of LFA algorithms reported so far, researchers have used forward and reverse sweep mechanism predominately. As leaf (terminal) node identification is a vital component to run LFA algorithm, while estimating the network branch currents during the reverse sweep, so work reported in this chapter mainly proposes an new LFA algorithm, which utilize the efficient scheme for leaf node identification proposed by Chaturvedi and Prasad[20].

This chapter is organized as follows. The mathematical modeling of load flow analysis is reported in section 3.2; a new load flow algorithms is covered in section 3.3. Performance analysis of proposed algorithm and comparison with earlier reported studies covered in section 3.4.

## 3.2 Mathematical Model of Radial Distribution Network

In RDNs, the large $R/X$ ratio causes problems in convergence of conventional load flow algorithms. For a balanced RDN, the network can be represented by an equivalent single-line diagram. The line shunt capacitances at distribution voltage level are very small and thus can be neglected. The simplified mathematical model of a section of a RDN is shown in Fig.3.1.

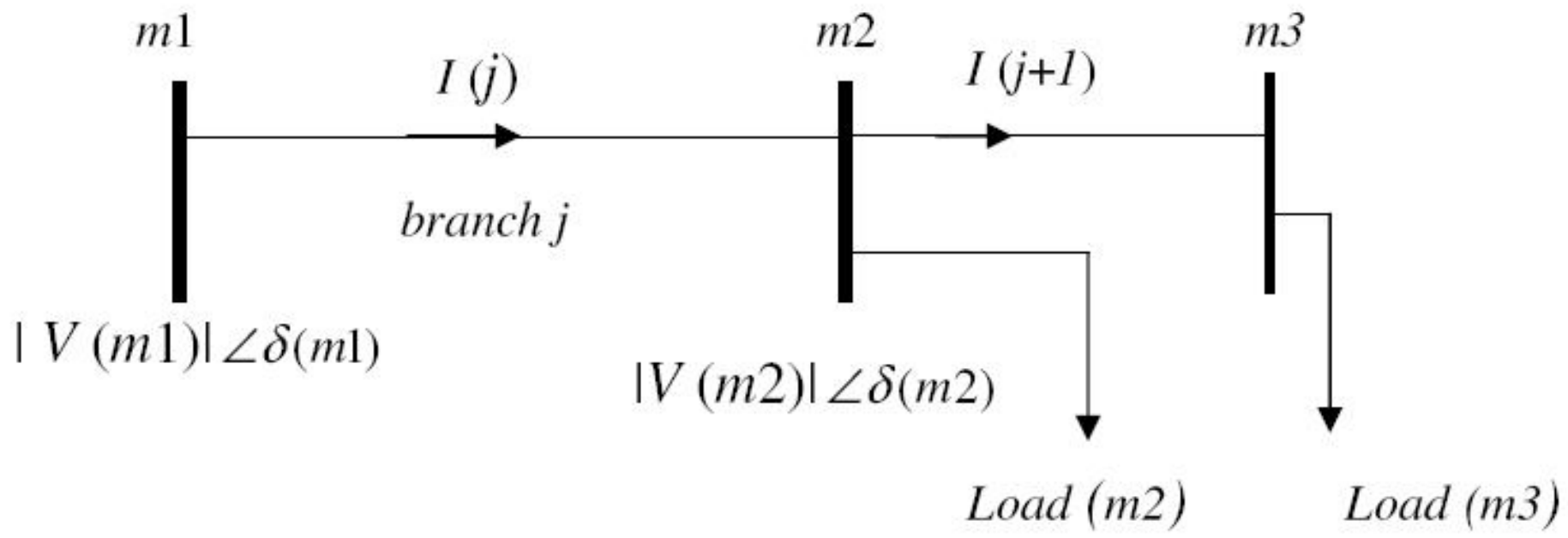

**Figure 3.1 : Electrical equivalent diagram of RDN for load flow study**

The complex power fed to node i can be represented by

$$S_i = V_i (LI)_i^* = P_i + jQ_i \tag{1}$$

$$(LI)_i = \left(\frac{S_i}{V_i}\right)^* = \frac{PL_i - jQL_i}{V_i^*} \tag{2}$$

$$= \frac{\sqrt{PL_i^2 + QL_i^2}}{|V_i|} \frac{\angle tan^{-1}(-QL_i / PL_i)}{\angle -\theta_{v_i}} \tag{3}$$

$$= |LI_i| \angle \theta_i$$

$$= |LI_i| cos\theta_i + j|LI_i| sin\theta_i$$

(4)

Where $|LI_i| = \frac{\left[PL_i^2 + QL_i^2\right]^{1/2}}{|V_i|}$ (5)

$$\theta_i = \theta_{v_i} - tan^{-1}\left(\frac{QL_i}{PL_i}\right) \tag{6}$$

**Branch current calculation:**

$$I_{br_j} = \sum_{i=1}^{n} |LI_i| \cos\theta_i + j\sum_{i=1}^{n} |LI_i| \sin\theta_i$$

$$= Re\left(I_{br_j}\right) + Im\left(I_{br_j}\right)$$

$I_{br\,j} = |I_{br\,j}| \angle I_{br\,j}$ where $|I_{br_j}| = \left[\left(Re\left(I_{br_j}\right)\right)^2 + \left(Im\left(I_{br_j}\right)\right)^2\right]^{1/2}$ (7)

and $\angle I_{br_j} = tan^{-1} \frac{Im\left(I_{br_j}\right)}{Re\left(I_{br_j}\right)}$ (8)

**Voltage calculation:**

$$V_r = V_s - I_{br}.Z_{br}$$

$$|V_r||\angle\theta V_r = |V_s|\angle\theta V_s - |I_{br}|\angle\theta I_{br}.|Z_{br}|.\angle\theta Z_{br}$$

$$= |V_s|\angle\theta V_s - |I_{br}|.|Z_{br}|\angle\theta I_{br} + \theta Z_{br}$$

$$= |V_s|\angle\theta V_s - |I_{br}|.|Z_{br}|\angle\phi \qquad (9)$$

On equating real and imaginary part equation (9) can be split as

$$|V_r|\cos\theta V_r = |V_s|\cos\theta V_s - |I_{br}|.|Z_{br}|\cos\phi \qquad (10)$$

and $|V_r|\sin\theta V_r = |V_s|\sin\theta V_s - |I_{br}|.|Z_{br}|\sin\phi$ (11)

Where $\phi = \theta I_{br} + \theta Z_{br} = \tan^{-1}\frac{Im(I_{br_j})}{Re(I_{br_j})} + \tan^{-1}\left(\frac{X_{br}}{R_{br}}\right)$ (12)

On squaring and adding equations (10) and (11) results in

$$|V_r|^2 = |V_s|^2 + |I_{br}|^2.|Z_{br}|^2 - 2|V_s|.|I_{br}|.$$
$$|Z_{br}|\{\cos\theta V_s \cos\phi + \sin\theta V_s \sin\phi\}$$

$$= |V_s|^2 + |I_{br}|^2.|Z_{br}|^2 - 2|V_s|.|I_{br}|.|Z_{br}|\cos(\theta V_s - \phi) \qquad (13)$$

On dividing equation (11) by equation (10), following expression results

$$\theta V_r = \tan^{-1}\left[\frac{|V_s|\sin\theta V_s - |I_{br}|.|Z_{br}|\sin\phi}{|V_s|\cos\theta V_s - |I_{br}|.|Z_{br}|\cos\phi}\right] \qquad (14)$$

Thus, once branch currents are computed, the node voltages are estimated using the above equations. Hence, the complexity of the solutions lies in the computation of branch currents. This paper presents a relatively simple and efficient procedure to identify the leaf node of a RDN and subsequently estimate the branch currents and node voltages. In a typical load flow study, without any prior knowledge, the following iterative procedure is followed.

*Step 1*: Read the system data and initially set all the node voltages to 1.0 p. u. (per unit) and branch currents to 0.

*Step 2*: Compute the currents for all the branches of the RDN.

*Step 3:* Update the node voltages using the computed branch currents.

*Step 4*: If the absolute value of the difference between the previous (iteration) and present (iteration) voltage at any node is more than some preset value (0.0001), then go to *Step 2* else *stop*.

Pseudo code for LFA algorithm is given in section 3.3. Once the convergence criterion is satisfied in LFA algorithm execution, the real and reactive power losses in a particular branch are computed as follows:

$$LP_j = |I_{br_j}|^2 . R_{br_j} \tag{15}$$

$$LQ_j = |I_{br_j}|^2 . X_{br_j} \tag{16}$$

The nomenclature of variables used in equations 1-16 are as follows.

$S_i$: Complex power fed at node *i*

$P_i$ :Real Power fed at node *i*

$Q_i$ :Reactive Power fed at node *i*

*NB*: Total number of nodes (*i* = 1,2, -------- *NB*)

*LN* : Total number of branches (*LN* =*NB*-1)

$PL_i$: Real power load at $i^{th}$ node

$QL_i$: Reactive power load at $i^{th}$ node

$|V_i|$: Voltage magnitude of $i^{th}$ node

$\theta_{v_i}$ : Voltage angle of $i^{th}$ node

$|LI_i|$: Load current magnitude at $i^{th}$ node

$\theta_i$ *: Load current angle at $i^{th}$ node*

$|I_{br_j}|$*: Current magnitude in branch j*

$\angle I_{br_j}$ : Current angle in branch $j$

$V_s$ : Sending node voltage

$V_r$ : Receiving node voltage

$IS(j)$: Sending end node of branch $j$

$IR(j)$: Receiving end node of branch $j$

$LP_j$ : Real power loss of branch $j$

$LQ_j$ : Reactive power loss of branch $j$

From the mathematical model of RDN, it is apparent that, the computation of voltage at each node and branch losses, can be evaluated only after finding the branch currents $I$ in all branches. The present work describes two different efficient LFA algorithms for branch currents estimation.

### 3.3 Pseudo-Code for LFA Algorithm

The step by step procedure for LFA implementation along with variables declaration is mentioned below:

Where N: Number of nodes in a given RDN.

$N_{b:}$ Number of branches in RDN and is notified as $N_b$= N-1.

NETWORK: It is multidimensional array that stores parameters of RDN.

$C_j$: column of NETWORK array as $C_1$ = branch number, $C_2$ = sending node number, $C_3$ = receiving node number, $C_4$ and $C_5$ = branch resistance & reactance, $C_6$ and $C_7$ = real & reactive load at the receiving end of the branch, respectively.

//**Step 1**: Reading the distribution system network data and storing in the array NETWORK.

V[i] = $V_{old}$[i] = 1.0 for i=2… N;     // Initial guess for the node voltages of RDS

I[j] = 0 for j= 1,…,$N_b$        //Initial guess for the branch currents; $N_b$= N-1

//**Step 2:** Using sub-routine LEAF [21]identify the total leaf nodes present in a RDN.

//**Step 3:** computation of branch current

```
fori=Nb  :  -1 : 1
        flag=1;                    //flag is a Boolean variable.//
        low=1;
        high=length(Leaf);
        while (low <= high)
                mid = floor((low + high) / 2);
                if (Leaf(mid) > NETWORK (i,3))
```

```
            high = mid - 1;
        elseif (Leaf(mid) < NETWORK (i,3))
            low = mid + 1;
        else
            flag=0   // it will set when the receiving end is also a leaf //
            break;
        end
    end
    if(flag==0)     // branch current of a branch that is connected to a leaf//
        I(NETWORK (i,1))=IL(NETWORK (i,3))
    end
    else
        top=0
        for j = 2 : Nb
            if (NETWORK (i,3)== NETWORK (j,2))
                top = top+1;
                St (top) = NETWORK (j, 1)
            end     // find the all branches for which the receiving node of
                    branch I is the sending node and push them into stack. //
        end
```

*// the following code estimates the current in branch i, for which receiving end node is a non-leaf node .//*

```
                while (top>=1)
                        BR = st (top);  // taking the top most element from the stack //
                        top = top-1;
                        I(NETWORK (i,1)) = I  (NETWORK (i,1)) + I(BR)
                end
                I(NETWORK (i,1)) = I(NETWORK (i,1)) + IL(NETWORK (i,3))
        end
end     //   end of  outer for loop. //
```

**//Step 4**: compute the new voltage for each node, i. e., $\mathbf{V}_{\mathbf{new}_i}$ where i = 2, 3…N

**//Step 5:** counter=0

```
fori=1 to N
        if (abs(Vnew_i)-abs(Vold_i))<= 0.0001
                counter = counter+1;
        end
end
if (counter==N )
        go to step 6
else
        go to step3
```

**//Step 6:** Return.

## 3.4 Complexity Analysis

The complexity of an algorithm is a function which gives the running time / or space in terms of the input size. In fact the time and space used by the algorithm are two main measures that decide the efficiency of any algorithm.

***(a) Time complexity*:** The amount of CPU time or the number of clock cycles required by an algorithm to solve a problem, which is expressed as a function of input data size is known as the time complexity of the problem. Sometime time complexity is measured by counting the number of key operations – in sorting and searching algorithms, for example, the number of comparison. That is because key operations are so defined that the time for the operations is much less than or at most proportional to the time for the key operations. The proposed LFA algorithm has been described in 6 steps. The step number. 1, 4, and 6 are representing constant time operation. At step 2 the proposed LFA algorithm uses one procedure LEAF[21], for finding the all leaf present in a given RDN. The time complexity of this procedure is $O(N)$. At step 3 algorithm estimates the branch currents corresponds to each leaf and non-leaf nodes of the given RDN. If a node is a leaf node then the branch current calculation for which it is the receiving node requires just one assignment operation, i.e. one unit time operation, hence, time complexity is O(1). If a node is not a leaf node, then the program scans through the sending node list to look for the branches for which it is a sending node. Again, the complexity is $O(N)$. Thus, for the worst case scenario, the time complexity of step 3 is $O(N*N)$. At step 5 algorithm checks the convergence condition and some time to get converge solution algorithm calls step 3 repeatedly inside step 5.Thus the step 5 will dominate the other remaining steps .so the overall time complexity of the proposed algorithm is of the $O(R* N^2)$,where R is number of iteration required to get the converged solutions.

***(b) Space complexity:***

The amount of memory required by an algorithm at run time to solve a problem, which is expressed as a function of input data size, is known as the space complexity of that problem. This algorithm uses array and a stack to store the data of size NB. There are no recursive procedures or dynamic memory allocations. Since the array and stack used are one dimensional, it is apparent that the space complexity is only $O(N)$.

# RESULT

In this thesis, the load flow algorithm proposed by Prasad et. al.[20] is amended on integrating sub-module for leaf node identification[26]. The modified LFA algorithm is tested on two different distribution networks, viz. 34-bus and 69-bus RDN. For both the networks voltage profile at all the network nodes (buses) so obtained from proposed LFA algorithm is same as one reported in [20] and is presented here in Table 1 for 69-bus RDN. However, uses of proposed scheme yield a remarkable saving in number of steps execution required to get converged load flow solutions. In general any n-node RDN consisting m leaf node and suppose r iteration are required for reaching steady state condition ,the proposed LFA algorithm requires approximately $3n + n^2 + r(n + n.m + n(n-m) + n)$ i.e. $O(n^2) + O(r*n^2)$ steps, while using LFA algorithm reported in [20],the number of steps attains a value $3n + n^2 + r(n + n^2 + n^2 + n)$ i.e. $O(n^2)+O(r*2n^2)$. Hence, there is significant saving in no. of steps execution as the value of m is very less as compared to the value of n.

**Table 4.1.** Comparison of load flow results for 69-node RDN between the proposed algorithm (♣) and the algorithm proposed in [20]

| Node Number | Voltage Magnitude (♣) (p.u.) | Voltage Magnitude [20] (p.u.) | Node Number | Voltage Magnitude (♣) (p.u.) | Voltage Magnitude [20] (p.u.) |
|---|---|---|---|---|---|
| 1 | 1.00000 | 1.00000 | 36 | 0.99992 | 0.99992 |
| 2 | 0.99997 | 0.99997 | 37 | 0.99975 | 0.99975 |
| 3 | 0.99993 | 0.99993 | 38 | 0.99959 | 0.99959 |
| 4 | 0.99984 | 0.99984 | 39 | 0.99954 | 0.99954 |
| 5 | 0.99902 | 0.99902 | 40 | 0.99954 | 0.99954 |
| 6 | 0.99008 | 0.99009 | 41 | 0.99884 | 0.99884 |
| 7 | 0.98079 | 0.98079 | 42 | 0.99855 | 0.99855 |
| 8 | 0.97857 | 0.97858 | 43 | 0.99851 | 0.99851 |
| 9 | 0.97744 | 0.97744 | 44 | 0.99850 | 0.99850 |
| 10 | 0.97243 | 0.97244 | 45 | 0.99841 | 0.99841 |
| 11 | 0.97131 | 0.97132 | 46 | 0.99840 | 0.99840 |
| 12 | 0.96814 | 0.96816 | 47 | 0.99979 | 0.99979 |
| 13 | 0.96521 | 0.96523 | 48 | 0.99854 | 0.99854 |
| 14 | 0.96231 | 0.96233 | 49 | 0.99469 | 0.99470 |
| 15 | 0.95943 | 0.95946 | 50 | 0.99415 | 0.99415 |
| 16 | 0.95890 | 0.95893 | 51 | 0.97854 | 0.97854 |
| 17 | 0.95802 | 0.95805 | 52 | 0.97853 | 0.97853 |
| 18 | 0.95801 | 0.95804 | 53 | 0.97465 | 0.97466 |
| 19 | 0.95754 | 0.95757 | 54 | 0.97141 | 0.97141 |
| 20 | 0.95724 | 0.95727 | 55 | 0.96693 | 0.96694 |
| 21 | 0.95676 | 0.95679 | 56 | 0.96256 | 0.96257 |
| 22 | 0.95675 | 0.95678 | 57 | 0.94004 | 0.94010 |
| 23 | 0.95668 | 0.95671 | 58 | 0.92894 | 0.92904 |
| 24 | 0.95652 | 0.95656 | 59 | 0.92464 | 0.92476 |
| 25 | 0.95635 | 0.95638 | 60 | 0.91958 | 0.91974 |
| 26 | 0.95628 | 0.95631 | 61 | 0.91217 | 0.91234 |
| 27 | 0.95626 | 0.95629 | 62 | 0.91188 | 0.91205 |
| 28 | 0.99993 | 0.99993 | 63 | 0.91149 | 0.91167 |
| 29 | 0.99985 | 0.99985 | 64 | 0.90958 | 0.90977 |
| 30 | 0.99973 | 0.99973 | 65 | 0.90901 | 0.90919 |
| 31 | 0.99971 | 0.99971 | 66 | 0.97125 | 0.97126 |
| 32 | 0.99961 | 0.99961 | 67 | 0.97125 | 0.97126 |
| 33 | 0.99935 | 0.99935 | 68 | 0.96781 | 0.96783 |
| 34 | 0.99901 | 0.99901 | 69 | 0.96781 | 0.96782 |
| 35 | 0.99895 | 0.99895 | | | |

# CONCLUSION AND FUTURE SCOPE

## 5.1 Conclusions :

On using the proposed Load Flow Analysis algorithm, considerable amount of saving can be achieved in number of steps execution, required to obtain steady state load flow solutions. Further, on account of algorithm’s complexity order; although, the overall complexity of the proposed Load Flow Analysis algorithm is approximately same as one proposed i[20]. As for static network topology, the uses of proposed scheme renders saving in time; it indicates that for the network optimization aspects, when network topology undergoes dynamic reconfiguration and thus, this process leads to number of distinct static network topologies. Obviously, for each of these static network topologies, saving in time required for load flow analysis for all these network topologies will be a phenomenal figure. There is significant saving in no. of steps execution as the value of m is very less as compared to the value of n.

## 5.2 Future scope :

Using graph theory concept and exploiting multi-cores architecture, the proposed method for load flow can be further investigated for obtaining more optimized solutions.

## APPENDIX-I

**Table A: System data for 69-bus radial distribution network ('*' denotes a tie-line)**

| Branch Number | Sending Bus | Receiving Bus | Resistance Ω | Reactance Ω | Nominal Load at Receiving Bus | | Maximum Line Capacity (kVA) |
|---|---|---|---|---|---|---|---|
| | | | | | P (kW) | Q (kVAr) | |
| 1 | 1 | 2 | 0.0005 | 0.0012 | 0.0 | 0.0 | 10761 |
| 2 | 2 | 3 | 0.0005 | 0.0012 | 0.0 | 0.0 | 10761 |
| 3 | 3 | 4 | 0.0015 | 0.0036 | 0.0 | 0.0 | 10761 |
| 4 | 4 | 5 | 0.0251 | 0.0294 | 0.0 | 0.0 | 5823 |
| 5 | 5 | 6 | 0.3660 | 0.1864 | 2.60 | 2.20 | 1899 |
| 6 | 6 | 7 | 0.3811 | 0.1941 | 40.40 | 30.00 | 1899 |
| 7 | 7 | 8 | 0.0922 | 0.0470 | 75.00 | 54.00 | 1899 |
| 8 | 8 | 9 | 0.0493 | 0.0251 | 30.00 | 22.00 | 1899 |
| 9 | 9 | 10 | 0.8190 | 0.2707 | 28.00 | 19.00 | 1455 |
| 10 | 10 | 11 | 0.1872 | 0.0619 | 145.00 | 104.00 | 1455 |
| 11 | 11 | 12 | 0.7114 | 0.2351 | 145.00 | 104.00 | 1455 |
| 12 | 12 | 13 | 1.0300 | 0.3400 | 8.00 | 5.00 | 1455 |
| 13 | 13 | 14 | 1.0440 | 0.3450 | 8.00 | 5.50 | 1455 |
| 14 | 14 | 15 | 1.0580 | 0.3496 | 0.0 | 0.0 | 1455 |
| 15 | 15 | 16 | 0.1966 | 0.0650 | 45.50 | 30.00 | 1455 |
| 16 | 16 | 17 | 0.3744 | 0.1238 | 60.00 | 35.00 | 1455 |
| 17 | 17 | 18 | 0.0047 | 0.0016 | 60.00 | 35.00 | 2200 |
| 18 | 18 | 19 | 0.3276 | 0.1083 | 0.0 | 0.0 | 1455 |
| 19 | 19 | 20 | 0.2106 | 0.0690 | 1.00 | 0.60 | 1455 |
| 20 | 20 | 21 | 0.3416 | 0.1129 | 114.00 | 81.00 | 1455 |
| 21 | 21 | 22 | 0.0140 | 0.0046 | 5.00 | 3.50 | 1455 |
| 22 | 22 | 23 | 0.1591 | 0.0526 | 0.0 | 0.0 | 1455 |
| 23 | 23 | 24 | 0.3463 | 0.1145 | 28.00 | 20.0 | 1455 |
| 24 | 24 | 25 | 0.7488 | 0.2475 | 0.0 | 0.0 | 1455 |
| 25 | 25 | 26 | 0.3089 | 0.1021 | 14.0 | 10.0 | 1455 |
| 26 | 26 | 27 | 0.1732 | 0.0572 | 14.0 | 10.0 | 1455 |
| 27 | 3 | 28 | 0.0044 | 0.0108 | 26.0 | 18.6 | 10761 |
| 28 | 28 | 29 | 0.0640 | 0.1565 | 26.0 | 18.6 | 10761 |
| 29 | 29 | 30 | 0.3978 | 0.1315 | 0.0 | 0.0 | 1455 |
| 30 | 30 | 31 | 0.0702 | 0.0232 | 0.0 | 0.0 | 1455 |
| 31 | 31 | 32 | 0.3510 | 0.1160 | 0.0 | 0.0 | 1455 |
| 32 | 32 | 33 | 0.8390 | 0.2816 | 14.0 | 10.0 | 2200 |
| 33 | 33 | 34 | 1.7080 | 0.5646 | 9.50 | 14.00 | 1455 |
| 34 | 34 | 35 | 1.4740 | 0.4873 | 6.00 | 4.00 | 1455 |
| 35 | 3 | 36 | 0.0044 | 0.0108 | 26.0 | 18.55 | 10761 |
| 36 | 36 | 37 | 0.0640 | 0.1565 | 26.0 | 18.55 | 10761 |
| 37 | 37 | 38 | 0.1053 | 0.1230 | 0.0 | 0.0 | 5823 |
| 38 | 38 | 39 | 0.0304 | 0.0355 | 24.0 | 17.00 | 5823 |
| 39 | 39 | 40 | 0.0018 | 0.0021 | 24.0 | 17.00 | 5823 |
| 40 | 40 | 41 | 0.7283 | 0.8509 | 1.20 | 1.0 | 5823 |
| 41 | 41 | 42 | 0.3100 | 0.3623 | 0.0 | 0.0 | 5823 |
| 42 | 42 | 43 | 0.0410 | 0.0478 | 6.0 | 4.30 | 5823 |
| 43 | 43 | 44 | 0.0092 | 0.0116 | 0.0 | 0.0 | 5823 |
| 44 | 44 | 45 | 0.1089 | 0.1373 | 39.22 | 26.30 | 5823 |
| 45 | 45 | 46 | 0.0009 | 0.0012 | 39.22 | 26.30 | 6709 |
| 46 | 4 | 47 | 0.0034 | 0.0084 | 0.00 | 0.0 | 10761 |
| 47 | 47 | 48 | 0.0851 | 0.2083 | 79.00 | 56.40 | 10761 |
| 48 | 48 | 49 | 0.2898 | 0.7091 | 384.70 | 274.50 | 10761 |
| 49 | 49 | 50 | 0.0822 | 0.2011 | 384.70 | 274.50 | 10761 |
| 50 | 8 | 51 | 0.0928 | 0.0473 | 40.50 | 28.30 | 1899 |
| 51 | 51 | 52 | 0.3319 | 0.1114 | 3.60 | 2.70 | 2200 |
| 52 | 52 | 53 | 0.1740 | 0.0886 | 4.35 | 3.50 | 1899 |
| 53 | 53 | 54 | 0.2030 | 0.1034 | 26.40 | 19.00 | 1899 |
| 54 | 54 | 55 | 0.2842 | 0.1447 | 24.00 | 17.20 | 1899 |
| 55 | 55 | 56 | 0.2813 | 0.1433 | 0.0 | 0.0 | 1899 |
| 56 | 56 | 57 | 1.5900 | 0.5337 | 0.0 | 0.0 | 2200 |
| 57 | 57 | 58 | 0.7837 | 0.2630 | 0.0 | 0.0 | 2200 |
| 58 | 58 | 59 | 0.3042 | 0.1006 | 100.0 | 72.0 | 1455 |
| 59 | 59 | 60 | 0.3861 | 0.1172 | 0.0 | 0.0 | 1455 |
| 60 | 60 | 61 | 0.5075 | 0.2585 | 1244.0 | 888.00 | 1899 |
| 61 | 61 | 62 | 0.0974 | 0.0496 | 32.0 | 23.00 | 1899 |
| 62 | 62 | 63 | 0.1450 | 0.0738 | 0.0 | 0.0 | 1899 |
| 63 | 63 | 64 | 0.7105 | 0.3619 | 227.0 | 162.00 | 1899 |
| 64 | 64 | 65 | 1.0410 | 0.5302 | 59.0 | 42.0 | 1899 |
| 65 | 11 | 66 | 0.2012 | 0.0611 | 18.0 | 13.0 | 1455 |
| 66 | 66 | 67 | 0.0047 | 0.0014 | 18.0 | 13.0 | 1455 |
| 67 | 12 | 68 | 0.7394 | 0.2444 | 28.0 | 20.0 | 1455 |
| 68 | 68 | 69 | 0.0047 | 0.0016 | 28.0 | 20.0 | 1455 |
| 69* | 11 | 43 | 0.5000 | 0.5000 | | | 566 |
| 70* | 13 | 21 | 0.5 | 0.5 | | | 566 |
| 71* | 15 | 46 | 1.0 | 1.0 | | | 400 |
| 72* | 50 | 59 | 2.0 | 2.0 | | | 283 |
| 73* | 27 | 65 | 1.0 | 1.0 | | | 400 |

**69 BUS RADIAL DISTRIBUTION NETWORK**

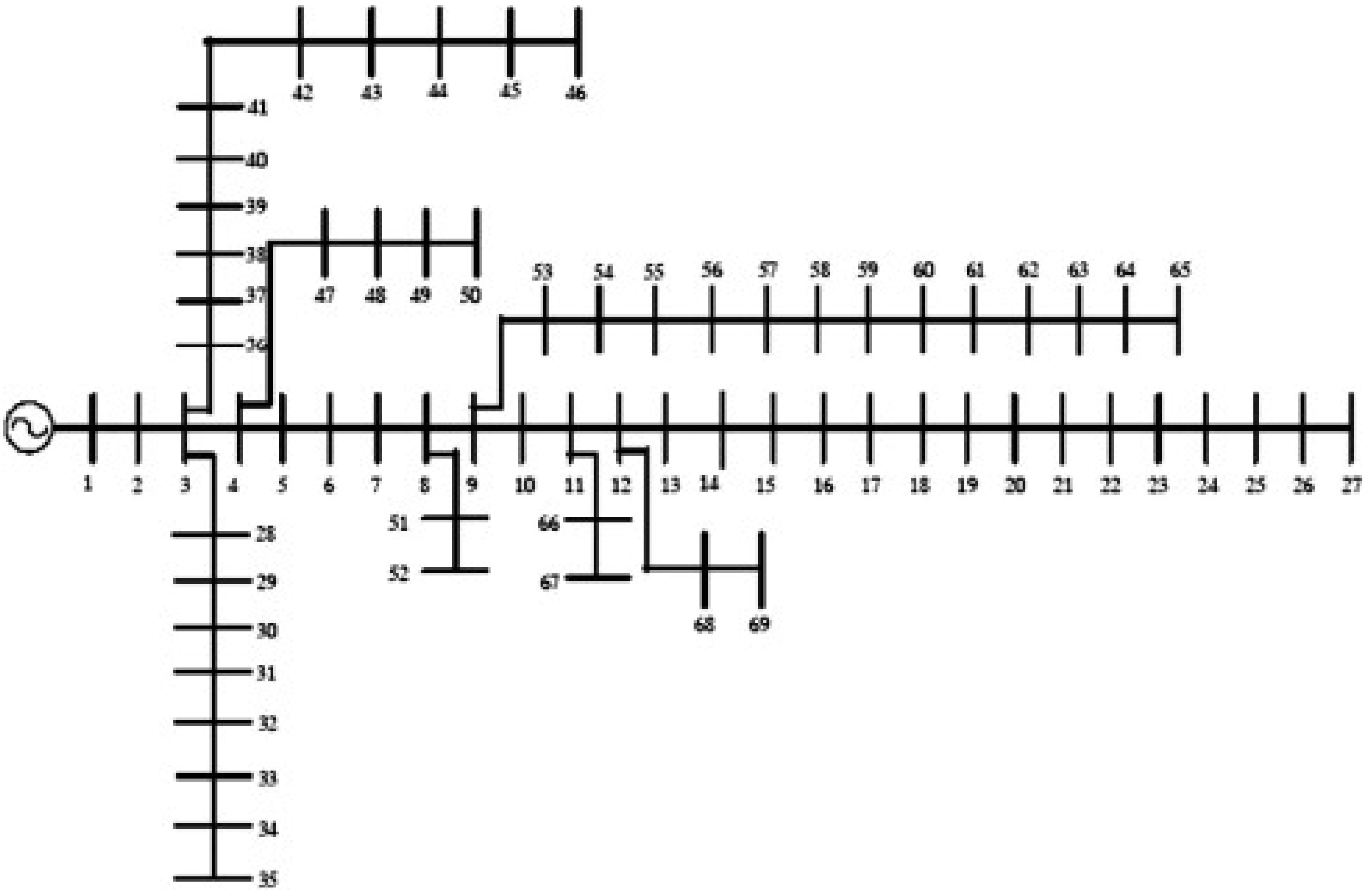

**69 – Bus Radial Distribution Network**

## APPENDIX-II

**Table B : System data for 33-bus radial distribution network**

| Branch Number | Sending Bus | Receiving Bus | Resistance Ω | Reactance Ω | Nominal Load at Receiving Bus | |
|---|---|---|---|---|---|---|
| | | | | | P (kW) | Q (kVAr) |
| 1 | 1 | 2 | 0.0922 | 0.047 | 100 | 60 |
| 2 | 2 | 3 | 0.493 | 0.2511 | 90 | 40 |
| 3 | 3 | 4 | 0.366 | 0.1864 | 120 | 80 |
| 4 | 4 | 5 | 0.3811 | 0.1941 | 60 | 30 |
| 5 | 5 | 6 | 0.819 | 0.707 | 60 | 20 |
| 6 | 6 | 7 | 0.1872 | 0.6188 | 200 | 100 |
| 7 | 7 | 8 | 0.7114 | 0.2351 | 200 | 100 |
| 8 | 8 | 9 | 1.03 | 0.74 | 60 | 20 |
| 9 | 9 | 10 | 1.044 | 0.74 | 60 | 20 |
| 10 | 10 | 11 | 0.1966 | 0.065 | 45 | 30 |
| 11 | 11 | 12 | 0.3744 | 0.1298 | 60 | 35 |
| 12 | 12 | 13 | 1.468 | 1.155 | 60 | 35 |
| 13 | 13 | 14 | 0.5416 | 0.7129 | 120 | 80 |
| 14 | 14 | 15 | 0.591 | 0.526 | 60 | 10 |
| 15 | 15 | 16 | 0.7463 | 0.545 | 60 | 20 |
| 16 | 16 | 17 | 1.289 | 1.721 | 60 | 20 |
| 17 | 17 | 18 | 0.732 | 0.574 | 90 | 40 |
| 18 | 2 | 19 | 0.164 | 0.1565 | 90 | 40 |
| 19 | 19 | 20 | 1.5042 | 1.3554 | 90 | 40 |
| 20 | 20 | 21 | 0.4095 | 0.4784 | 90 | 40 |
| 21 | 21 | 22 | 0.7089 | 0.9373 | 90 | 40 |
| 22 | 3 | 23 | 0.4512 | 0.3083 | 90 | 50 |
| 23 | 23 | 24 | 0.898 | 0.7091 | 420 | 200 |
| 24 | 24 | 25 | 0.896 | 0.7011 | 420 | 200 |
| 25 | 6 | 26 | 0.203 | 0.1034 | 60 | 25 |
| 26 | 26 | 27 | 0.2842 | 0.1447 | 60 | 25 |
| 27 | 27 | 28 | 1.059 | 0.9337 | 60 | 20 |
| 28 | 28 | 29 | 0.8042 | 0.7006 | 120 | 70 |
| 29 | 29 | 30 | 0.5075 | 0.2585 | 200 | 600 |
| 30 | 30 | 31 | 0.9744 | 0.963 | 150 | 70 |
| 31 | 31 | 32 | 0.3105 | 0.3619 | 210 | 100 |
| 32 | 32 | 33 | 0.341 | 0.5302 | 60 | 40 |

**33 BUS RADIAL DISTRIBUTION NETWORK**

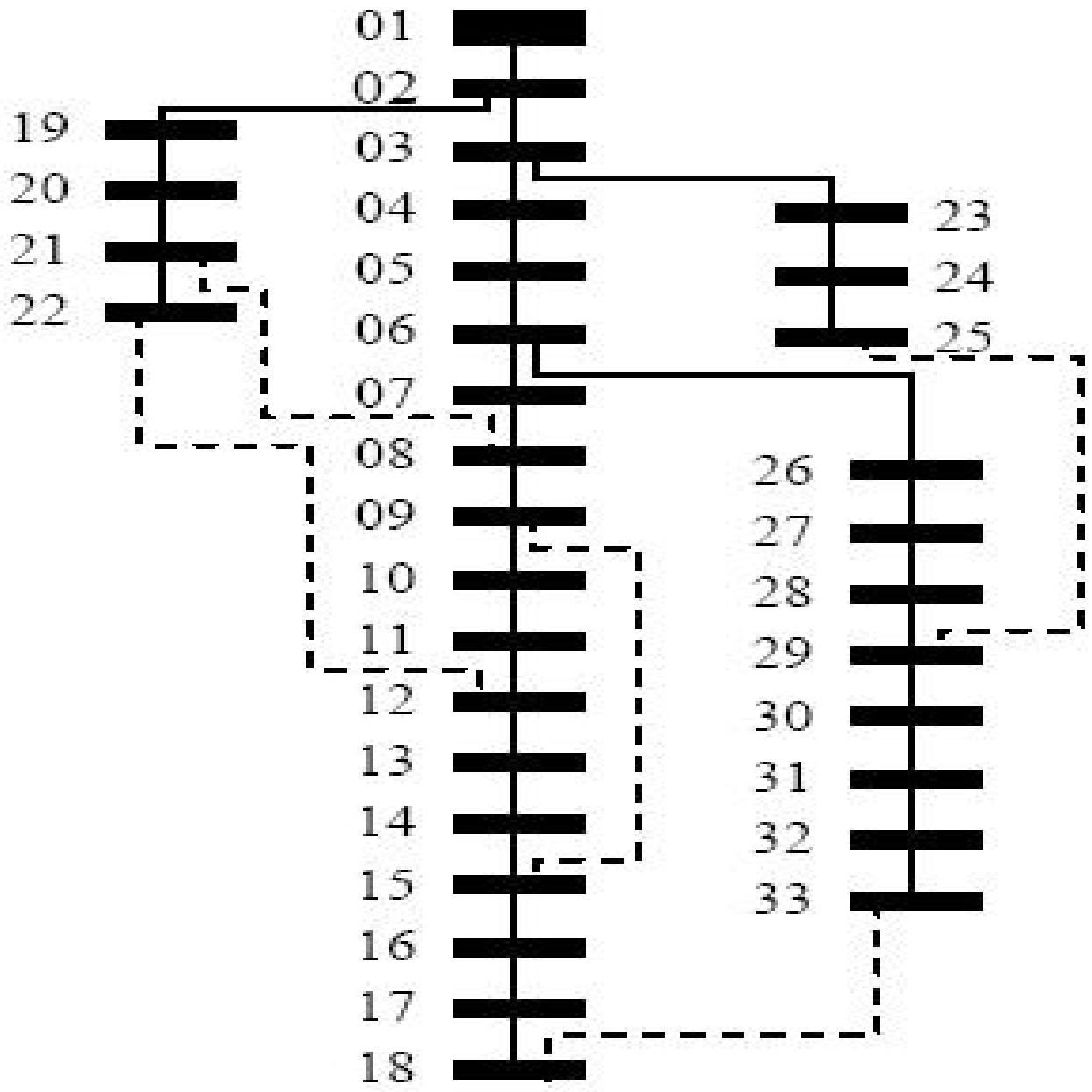

**33 - Bus Radial Distribution Network**